\documentclass[pra,twocolumn,showpacs,preprintnumbers,amsmath,amssymb]{revtex4}
\usepackage{graphicx}
\usepackage{dcolumn}
\usepackage{bm}
\begin{document}
\title{Lifetime of weakly-bound dimers of ultracold metastable helium studied by photoassociation}
\author{S. Moal, M. Portier, N. Zahzam, M. Leduc}
%
%
\affiliation{\'{E}cole Normale Sup\'{e}rieure, Laboratoire Kastler Brossel, 24 rue Lhomond, 75231 Paris Cedex 05, France.}
\date{\today}
%
\begin{abstract}
We describe two-photon photoassociation (PA) experiments operated in an ultracold gas of metastable $^4$He$^*$ atoms in the $2^3S_1$ state. Atom-molecule dark resonances as well as Raman signals provide information on the exotic molecule in the least bound vibrational $J=2$, $v=14$ state of the $^5\Sigma_g^+$ interaction potential of two spin-polarized metastable atoms. The physical origin of the various two-photon PA signals is first discussed. Their linewidths are interpreted in term of processes limiting the lifetime $\tau$ of the exotic molecule. A value of $\tau=1.4\pm0.3~\mu$s is found, which significantly differs from two recent calculations of the Penning ionization rate induced by spin-dipole coupling \cite{Whitt06,Gora}. The limitation of $\tau$ by atom-molecule inelastic collisions is investigated and ruled out. 
\end{abstract}

\maketitle

%
\section{Introduction}
\label{intro}

Helium atoms $^4$He$^*$ in the metastable $2^3S_1$ state  are long lived (2 hours lifetime) and can be cooled to very low temperature, down to Bose Einstein condensation, in magnetic traps in which they are spin polarized. The spin polarization is a drastic way to inhibit the inelastic collisions which otherwise would quickly destroy the atomic gas by Penning ionization. Such inhibition has been predicted by several authors \cite{Goran,Fedichev,Venturi} and results from the conservation of the electronic spin during the collision between two $2^3S_1$ spin polarized helium atoms. This inhibition has a limit set by spin relaxation processes occuring between a pair of colliding atoms. An inhibition of four order of magnitude, has been evaluated and confirmed by the possibility to achieve Bose Einstein condensation \cite{Robert,Pereira}. 

An opportunity to study Penning ionization between spin-polarized $2^3S_1$ atoms is offered by photoassociation experiments, initially set for high accuracy spectroscopy. In a recent experiment \cite{2photon} based on two-photon photoassociation, our group studied exotic molecules in which two spin-polarized metastable helium atoms are bound. Thes molecules exhibit a relatively long lifetime, in the microsecond range or above. They are in the least bound state of the $^5\Sigma^+_g$ interaction potential between the two atoms and their extension is quite large (of order 4~nm). Measuring the energy of this weakly bound molecular state using atom-molecule dark resonances allowed a very precise determination of the s-wave scattering length of the two polarized metastable atoms \cite{2photon}. 

This article is meant to present the latest results that we obtained about these exotic helium molecules. It first provides explanations about the physical origin of the two-photon photoassociation signals reported in reference \cite{2photon}. Fits are given for the dark resonance signals resulting from the free-bound-bound transition scheme. They use the formalism of references \cite {Bohn96,Bohn99} and incorporate the thermal averaging over the atom velocity. Then the linewidths of the two-photon signals are studied in detail, from which information about the lifetime of the exotic molecule can be derived. These experimental results are finally discussed in regard of two recent theoretical estimates of the spin-relaxation induced Penning ionization rate for the two polarized atoms forming an helium dimer \cite{Whitt06,Gora}. The possible influence of atom-molecule inelastic collisions is also discussed.

\begin{figure}[htbp]
\resizebox{1.0\columnwidth}{!}{%
  \includegraphics{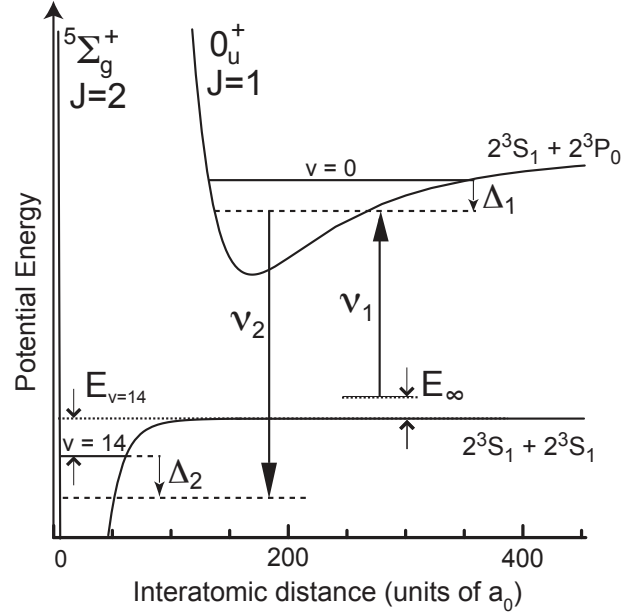}
}
\caption{Levels involved in the two-photon photoassociation experiment. $^5\Sigma^+_g$ is the interaction potential between two spin-polarized $2^3S_1$ helium atoms, $v=14$ being the least-bound vibrational state in this potential. Laser $L_1$ at frequency $\nu_1$ operates on the free-bound transition with a detuning $\Delta_1$, and laser $L_2$ at frequency $\nu_2$ drives the bound-bound transition with a detuning $\Delta_2$ from the two-photon transition. Energies are not to scale.
}
\label{fig:pot}
\end{figure}

%
\section{Obtention of two-photon photoassociation signals}
\label{PAsignal}

\subsection{Experimental set-up}

The two-photon scheme of the photoassociation experiment is shown in Fig.~\ref{fig:pot}. A pair of atoms in the metastable $2^3S_1$ state enter a collisional channel with a kinetic energy $E_{\infty}$. They are spin-polarized and interact through the $^5\Sigma^+_g$ potential.

The spin-polarized ultracold atomic $^4$He$^*$ cloud is confined in a three-coils magnetic trap of the Ioff\'e-Pritchard type. The bias magnetic field is typically 3~G. The cloud is kept at a temperature usually close to 4~$\mu$K and which can be varied in the range between 2 and 10 $\mu$K.  The density is of the  order of 10$^{13}$~cm$^{-3}$. The cloud is illuminated along the magnetic field axis by a photoassociation (PA) pulse of duration $\tau_{\mathrm{PA}}$\ ranging from 0.2 to 100~ms. After this PA pulse which produces a heating of the atomic cloud, a delay $\tau_{th}$ of typically 400~ms is introduced before one switches off the magnetic trap, in order to have a complete thermalization. A ballistic expansion follows before the detection takes place. The detection is optical, based on the absorption of a resonant probe beam by the atomic cloud. One registers simultaneously the number of atoms, the optical density at the center of the cloud (peak density) and the temperature raise, exploiting the calorimetric method described in \cite{Leonardexp,Kim}. The increase of temperature is due to the desexcitation of the molecules formed by the absorption of a photon into a pair of atoms taking away the binding energy of the molecular state and colliding with the cold trapped atoms. 

The PA laser is produced by a DBR (Distributed Bragg Reflector) diode laser at a wavelength of 1083~nm, operating in an external cavity (linewidth around 300~kHz). Its frequency is stabilized by an heterodyne technique as described in~\cite{Kim}. It is tuned close to the PA transition reaching the bound state $J=1$ $v=0$ in the $0_u^+$ potential, 1.4~GHz in the red of the $2^3S_1-2^3P_0$ atomic transition. The PA beam is amplified using an Ytterbium-doped fiber amplifier of 1~W power. Two gaussian laser beams are generated from this light source at two different frequencies $\nu_1$ and $\nu_2$ (see Fig.~\ref{fig:pot}), independently controlled by two acousto-optic modulators (AOM). They are superimposed and simultaneously focussed through the same optical fiber onto the atomic cloud after being $\sigma$- polarized. The beam waist is about 220~$\mu$m at the cloud position. The AOM devices are used to monitor the interruption of the laser beams by switching off the radio-frequency power supply. When it is off, we checked that no residual PA light perturbs the cloud. In each situation the pulse duration $\tau_{\mathrm{PA}}$\ is the same for the two beams. It is adjusted according to the PA beam power so as to approximately keep a constant amplitude of the signals.

\subsection{Autler-Townes and Raman spectra}
\label{signal}

Different spectroscopic signals can be recorded with the two PA beams when varying their frequencies.
Most commonly we use the first laser as a probe beam for plotting the spectra. We vary its frequency $\nu_1$ scanning  across the free-bound transition between the incoming two atoms and the $v=0$ state in the $0_u^+$ potential (see Fig.~\ref{fig:pot}). The second photon is at a fixed frequency $\nu_2$ close to the bound-bound transition between the two bound states $v=14$ in the $^5\Sigma_g^+$ and $v=0$ in the $0_u^+$ potential (see Fig.~\ref{fig:pot}). The definitions of the detunings $\Delta_1$ and $\Delta_2$ are the same as in \cite{Bohn99} (see Fig.~\ref{fig:pot}). Note that here $\Delta_2$ is not the detuning of the photon 2 from the bound-bound transition. When recording experimental spectra, we define $\Delta_1'$, which is the detuning of the first photon from the atomic transition $2^3S_1 \rightarrow 2^3P_0$. 

\begin{figure}[!]
  \includegraphics[angle=-90,scale=0.25]{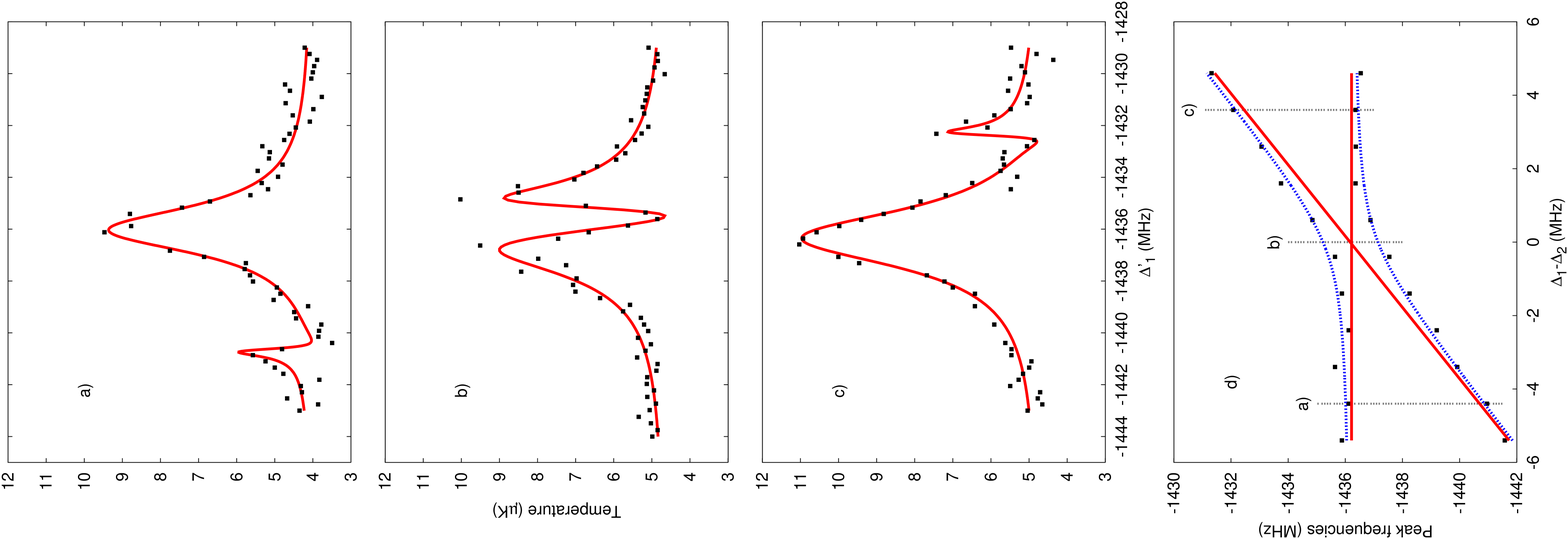}

\caption{Two-colour photoassociation spectra for various detunings of $\nu_2$ from the bound-bound transition. The two molecular bound states are dressed with the light field of laser $L_2$, forming a doublet probed by laser $L_1$. The temperature of the cloud is plotted as a function of frequency detuning $\Delta'_1$ of laser $L_1$ from the atomic transition. The intensities are $I_1 = 13$ mW/cm$^2$, $I_2 = 26$ mW/cm$^2$ and the PA pulse duration $\tau_{\mathrm{PA}} = 500$ $\mu$s. Raman configuration is obtained in cases a) ($\Delta_1-\Delta_2<0$) and c) ($\Delta_1-\Delta_2>0$). Case b) corresponds to an Autler-Townes configuration ($\Delta_1-\Delta_2=0$). The solid lines are fit curves based on the theoretical model described in \cite{Bohn96,Bohn99}. Figure d) shows the frequency position of the two peaks. Coupling of the two molecular states by laser $L_2$ results in an anticrossing.
}
\label{fig:at}
\end{figure}

\begin{figure}[!]
  \includegraphics[angle=-90,scale=0.25]{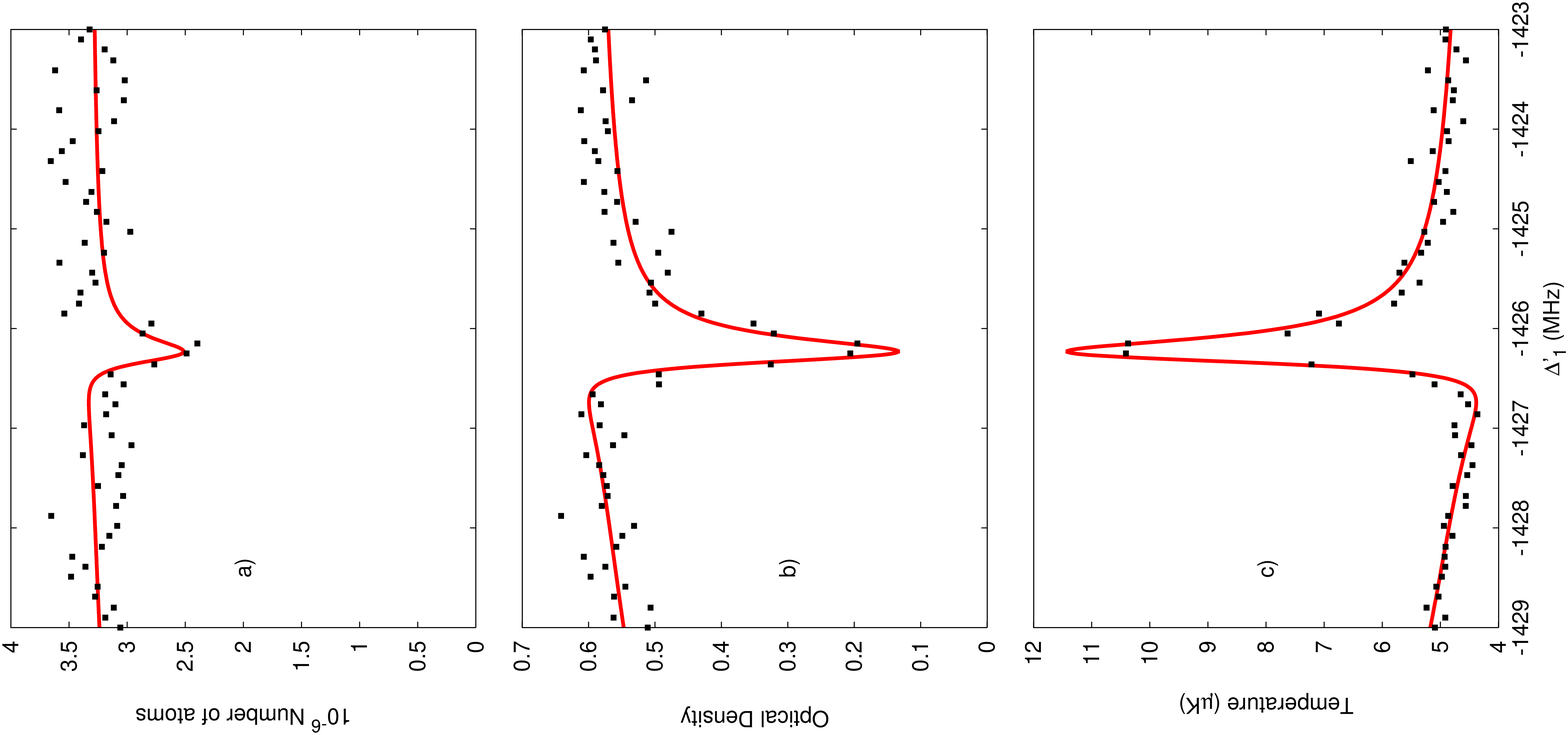}
\caption{Zoom on the Raman peak obtained with the experimental parameters $\Delta_1 - \Delta_2 = 10$ MHz, $I_1 = 260$ mW/cm$^2$, $I_2 = 65$ mW/cm$^2$, and $\tau_{\mathrm{PA}} = 200$ $\mu$s. 
After the PA laser pulse and thermalization during a time $\tau_{th}=400$~ms, the remaining atoms are detected optically. The atom number a), peak optical density b), and temperature c) are recorded.
}
\label{fig:3ramans}
\end{figure}

We study here the case where the detuning $\Delta_1-\Delta_2$ of the second photon from the bound-bound transition, and/or the Rabi frequency $\Omega$ associated to the bound-bound transition are larger than the width of the electronically excited state $\gamma_1$ ($\Omega,|\Delta_1-\Delta_2|\gtrsim\gamma_1/2\pi$). The application of the laser results in a splitting of the line associated to the molecular states that it couples \cite{API}. The PA signal recorded when varying $\Delta'_1$ shows two peaks (see Fig.~\ref{fig:at}).
If the second photon is exactly at resonance ($\Delta_2 = \Delta_1$), the two peaks are identical and the distance between them depends on the intensity of the second photon: it is the Autler-Townes configuration (see Fig.~\ref{fig:at}b). If the second photon is not exactly at resonance, the two  peaks are no longer identical. There is a main peak approximately at the same position as without the second photon, with a rather similar shape. Another smaller peak shows up: it is the Raman signal exhibiting a dissymetrical shape characterizing a Fano profile \cite{Fano61}. The position of the Raman peak is red detuned as compared to the main peak when the second photon is in the red of the bound-bound transition ($\Delta_1-\Delta_2<0$, see Fig.~\ref{fig:at}a) and blue detuned in the opposite case ($\Delta_1-\Delta_2>0$, see Fig.~\ref{fig:at}c). 

One can pass continuously from the Autler-Townes to the Raman configuration by scanning the difference $\Delta_1 - \Delta_2$ from zero to non-zero values. The positions of the two peaks vary as two hyperbolas as a function of $\Delta_1 - \Delta_2$ (see Fig.~\ref{fig:at}d). The two asymptotes represent the energies of the non-coupled states: the first one is that of the molecular excited state $v=0$ (horizontal asymptote of Fig.~\ref{fig:at}d), and the other one is that of the $v=14$ state plus the energy of the second photon (diagonal asymptote) (see \cite{API}). The effect of the coupling due to the second photon results in an anti-crossing as shown in Fig.~\ref{fig:at}d. Different positions of signals such as those shown in Fig.~\ref{fig:at}a,b and c are reported in Fig.~\ref{fig:at}d as a function of $\Delta_1 -\Delta_2$. The positions of the symmetrical Autler-Townes signals appear for $\Delta_1 - \Delta_2$ =0, whereas, for negative or positive values of $\Delta_1 - \Delta_2$, the positions of the Raman peaks appear on the hyperbolas close to the diagonal asymptote. 

We studied the Raman peaks in more detail, as we focus on the linewidth of these signals in the last section of this article. Fig.~\ref{fig:3ramans} shows a zoom of Raman signals such as those in Fig.~\ref{fig:at}c, here taken at a detuning $\Delta_1 -\Delta_2 = 10$~MHz. Three different signals are simultaneously recorded: the drop of the total number of atoms, the peak optical density and the temperature raise. The parameters have been chosen to optimize the temperature signal: the thermalization time $\tau_{th}$ after the PA pulse is sufficiently long ($\tau_{th}$ = 400~ms) to compensate for the short duration of the PA pulse ($\tau_{\mathrm{PA}} = 200$ $\mu$s). In Fig.~\ref{fig:3ramans} we observe a clear signal on the temperature raise and on the optical density, however, the number of atoms is not changing much (see Fig.~\ref{fig:3ramans}-a), contrary to the case of pure Autler-Townes or dark resonances when $\Delta_1-\Delta_2=0$ (see section \ref{sec:AMDRS}). The reason is that the number of excited molecules in the $v=0$ state is smaller when a large detuning $\Delta_1-\Delta_2$ is used for recording Raman signals: the number of metastable atoms is thus less modified when one scans through the resonance.

\subsection{Atom-molecule dark resonance signals}
\label{sec:AMDRS}
\label{darkres}

\begin{figure*}
 \includegraphics[angle=-90,scale=0.25]{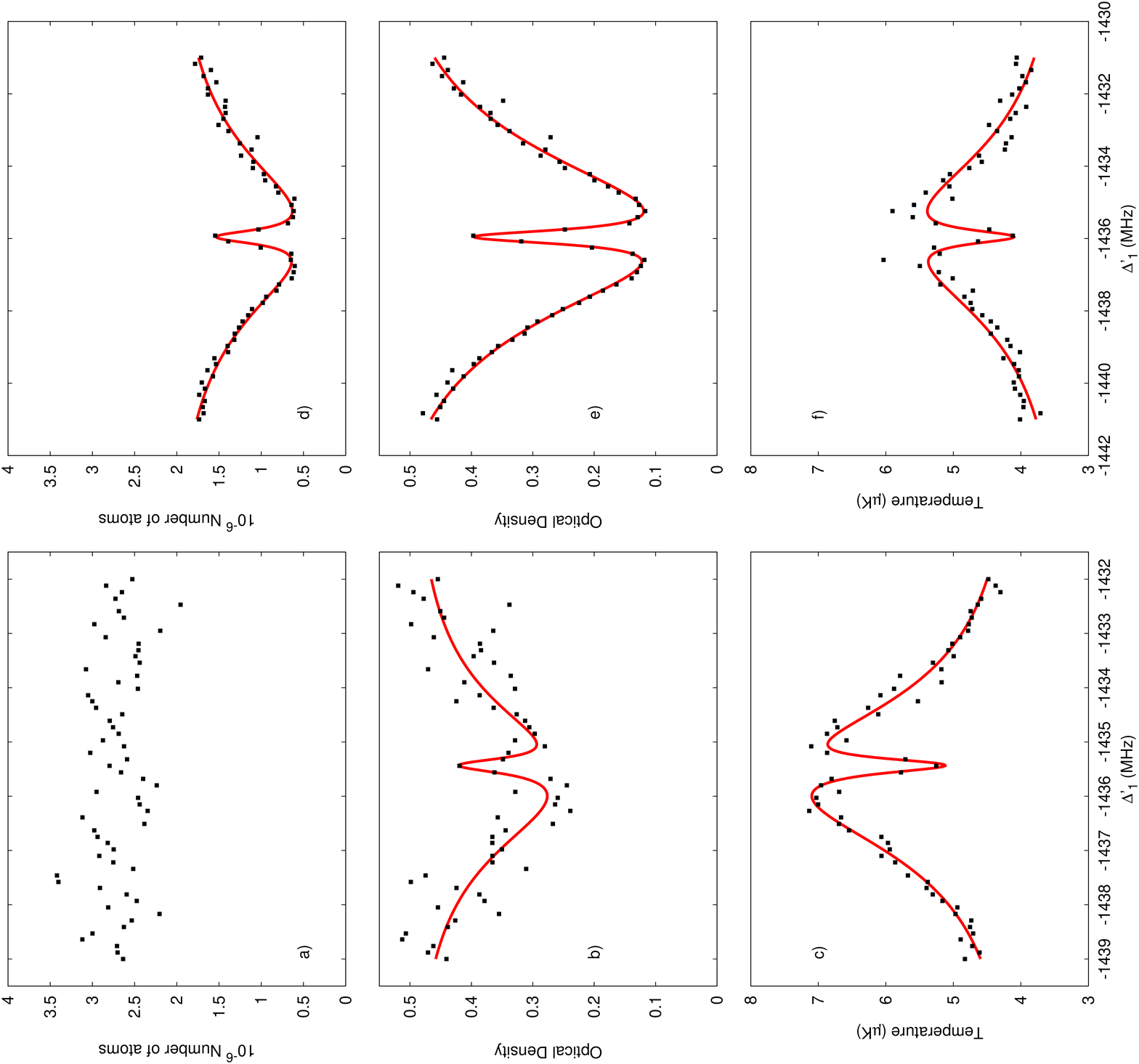}
\caption{Dark resonances in two-color photoassociation. The different spectra are obtained for $\Delta_1 - \Delta_2 = 0$ MHz. a), b), c) correspond to the experimental parameters $I_1 = 13$ mW/cm$^2$, $I_2 = 4$ mW/cm$^2$, $\tau_{\mathrm{PA}} = 500$ $\mu$s, $\tau_{th}$ = $400 ms$; d) e), f) correspond to $I_1 = I_2 = 13$ mW/cm$^2$, $\tau_{\mathrm{PA}} = 10$ ms, $\tau_{th}=0$ ms. After the PA laser pulses and further thermalization, the remaining atoms are detected optically. The atom number, peak optical density and temperature raise are recorded.
}
\label{fig:3signaux}
\end{figure*} 
When the second photon is set on resonance with the bound-bound transition ($\Delta_1 - \Delta_2 = 0$), and the intensity of the second laser beam is strongly decreased ($\Omega\lesssim\gamma_1$), one reaches a situation where the Autler-Townes splitting is replaced by a very narrow dip appearing at the center of a unique PA signal, as displayed in Fig.~\ref{fig:3signaux}. Such signals have been studied in \cite{2photon} and interpreted as dark resonances occuring between the pair of free atoms and the final $v=14$ bound molecular state. We are here in the exact Autler-Townes configuration as in Fig.~\ref{fig:at}b ($\Delta_2 = \Delta_1$) but with much lower intensity for the second photon $\nu_2$. The dark resonance results from the destructive interference between the two transition amplitudes for the two photons $\nu_1$ and $\nu_2$. Under these conditions, the width of the central dip is related to the lifetime of the $v=14$ state. The photoassociation should be totally inhibited if this lifetime was infinitly short, which is not the case. Note that the atom-molecule dark resonance condition is $\Delta_2=0$. It is studied for $\Delta_2=\Delta_1=0$, where the inhibition of the excitation of the $0_u^+$ molecular state is the most visible. It can also be observed for $\Delta_2=0$ and $\Delta_1\neq0$ on one side of the Raman peak.

Fig.~\ref{fig:3signaux} illustrates two different and extreme experimental conditions that we use to observe the dark resonance, according to the nature of the signal that we want to optimize. The first set of experimental parameters are chosen to optimize the temperature raise (Fig.~\ref{fig:3signaux}-c). For this, a large thermalization time $\tau_{th}$ (typically of 400~ms) favours the effect of heating due to the dissociation of the molecules in the $v=0$ state into pairs of fast atoms. In this case, one does not need a long PA pulse ($\tau_{\mathrm{PA}} = 500$ $\mu$s) nor large laser intensities. Thus the number of excited molecules is not sufficient to induce a significant loss of atoms (Fig.~\ref{fig:3signaux}-a). Note that the uncertainty on the number of atoms with these parameters is the worst and of the order of 30\%.
On the other hand, the second set of experimental parameters is chosen to optimize the atom loss (Fig.~\ref{fig:3signaux}-d), to the detriment of the temperature raise. Here a long PA pulse duration is chosen (alternatively a large intensity) for the first photon ($\tau_{\mathrm{PA}} = 10$ ms), namely 20 times larger than in the previous case, in order to increase the number of excited molecules and thus the metastable atom loss. However the thermalization time is here zero to avoid too large heating, which would blow up the cloud after the time of flight expansion.

\section{Theoretical analysis}
\label{theory}

\subsection{Formalism}
\label{formalism}

We use the theory of Bohn and Julienne \cite{Bohn96,Bohn99} to fit our data. The pair of two incoming metastable atoms is the initial state labelled $0$, the other two molecular states of the lambda transition scheme are $v=0$ and $v=14$ in the $0_u^+$ and $\mathrm{^5\Sigma^+_g}$ potentials respectively. The two decay amplitudes for the $0$ to $v=0$ and $0$ to $v=14$ transitions are described by the S-matrix elements $S_{0,v=0}$ and $S_{0,v=14}$, which can be written as: 
\begin{equation}
\label{eq:s012}
\left\{
\begin{array}{rl}
& |S_{0,v=0}|^2 =\displaystyle \frac{n(\Gamma,\Omega,\gamma_1,\gamma_2,\Delta_1,\Delta_2,E_\infty)}{d(\Gamma,\Omega,\gamma_1,\gamma_2,\Delta_1,\Delta_2,E_\infty)} \\
& \\
& |S_{0,v=14}|^2  =\displaystyle \frac{16 \Gamma \gamma_2 \Omega^2}{d(\Gamma,\Omega,\gamma_1,\gamma_2,\Delta_1,\Delta_2,E_\infty)}
\end{array}
\right. \, ,
\end{equation}
with
\begin{eqnarray}
\lefteqn{n(\Gamma,\Omega,\gamma_1,\gamma_2,\Delta_1,\Delta_2,E_\infty)=} \nonumber \\
&& 16 \Gamma \gamma_1 \left[ (E_\infty-\Delta_2)^2 + {\left( \frac{\gamma_2}{2} \right)}^2 \right] \nonumber \\
&& \nonumber \\
\lefteqn{d(\Gamma,\Omega,\gamma_1,\gamma_2,\Delta_1,\Delta_2,E_\infty)  =} \nonumber \\
&& 16 \Omega^4 -8\Omega^2 \left[ -\gamma_2 (\Gamma+\gamma_1) +4(E_\infty -\Delta_1) (E_\infty -\Delta_2) \right] \nonumber \\
&& + \left[ (\Gamma+\gamma_1)^2 +4(E_\infty -\Delta_1)^2 \right] \left[ \gamma_2^2 + 4(E_\infty -\Delta_2)^2 \right], \nonumber
\end{eqnarray}
where $\Gamma$ is the transition rate between the continuum state and $v=0$, $\Omega$ is the Rabi frequency for the transition between $v=0$ and $v=14$. $\gamma_1$ and $\gamma_2$ are the widths of, respectively, the $v=0$ and $v=14$ states ,$\Delta_{1,2}$ and $E_\infty$ are defined in Fig.~\ref{fig:pot}. $\gamma_1 /2 \pi \approx 3$~MHz according to \cite{Leonardexp}. $\Omega$ can be measured from the Autler-Townes spectra (see section~\ref{sec:COTRFOTSP}). $\Gamma$ is much smaller than $\gamma_1$ given the small intensity of the first laser ($\Gamma/2\pi$ is of the order of 1~kHz in the dark resonance experiments). The only adjustable parameter for the fit of the PA signals is $\gamma_2$, related to the lifetime of the $v=14$ state which we want to determine.

\subsection{Thermal averaging}
\label{thav}

In the case where photoassociation processes induce small atom losses, the PA signals are proportionnal to the rate coefficient $K_p$ of the inelastic processes yielding to the products of the light assisted collision between the two ground state $2^3S_1$ atoms. For a two-photon PA experiment involving a cloud of atoms characterized by a Maxwell-Boltzmann energy distribution, $K_p$ can be expressed as \cite{Napolitano}:
\begin{eqnarray}
\label{eq:thav}
\lefteqn{K_p(T,\Gamma,\Omega,\gamma_1,\gamma_2,\Delta_1,\Delta_2,E_\infty)
=\frac{h^2}{(2\pi \mu k_BT)^{3/2}} \times } \nonumber \\
&& \int_{0}^{\infty} \!\! {\left|S_{0,v=0}~\!\! \left(\! \Gamma,\Omega,\gamma_{1,2},\Delta_{1,2},\epsilon/k_B T \! \right) \! \right|}^2 e^{-\epsilon / k_BT} d\epsilon 
\end{eqnarray}
($\mu$ is the reduced mass and $k_B$ is the Boltzmann constant). 

One obtains $K_p$ as a function of $\gamma_2$, $\Omega$ and the temperature $T$ of the cloud. One measures $T$ by performing a time of flight expansion of the cloud. After the determination of  $\Omega$ using the Autler-Townes splitting (section~\ref{sec:COTRFOTSP}), we can fit our experimental spectra with formula~(\ref{eq:thav}), adding an offset and an appropriate multiplication factor to match the amplitude.

\subsection{Calibration of the Rabi frequency of the second photon}
\label{sec:COTRFOTSP}
\label{calibration}

\begin{figure}

  \includegraphics[angle=-90,scale=0.25]{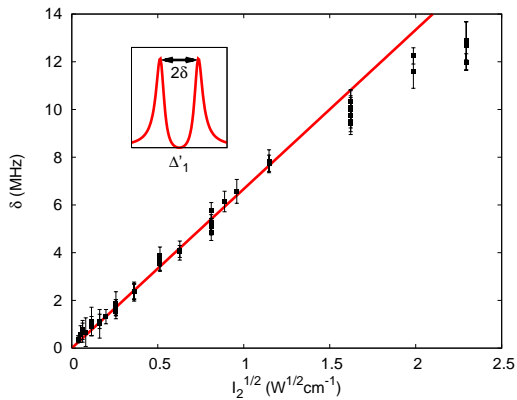}

\caption{
Rabi frequency calibration of the second photon. In the Autler-Townes configuration ($\Delta_1=\Delta_2$ with $I_1 = 260$ mW/cm$^2$, $\tau_{\mathrm{PA}} = 100$ $\mu$s), the half-splitting between the two peaks is plotted versus the square root of the second photon intensity $I_2$. The inset shows a typical Autler-Townes spectrum where two peaks are separated by $2\delta$.
}
\label{fig:atintensity}
\end{figure}

The distance $2h\delta$ between the two peaks in Fig.~\ref{fig:at} is given by equation (\ref{eq:hyperbole}), which is valid as far as $\delta\gtrsim\gamma_1/2\pi$ \cite{API}: 
\begin{equation}
\label{eq:hyperbole}
2 h \delta = \hbar \sqrt{{4 \pi^2 (\Delta_1 -\Delta_2)}^2 + 4 \Omega^2} \, .
\end{equation}
For the Autler-Townes signal ($\Delta_2=\Delta_1$), this reduces to $\delta=\Omega$. We have plotted the half splitting $\delta$ as a function of the square root of the second photon intensity $I_2$, as shown in Fig.~\ref{fig:atintensity}. By a linear fit, we obtain a calibration of the Rabi frequency : 
$$\Omega /\sqrt{I_2} =(6.67 \pm 1.00)~\mathrm{MHz.cm.W^{-1/2}}$$
The linear fit can be applied up to $\sqrt{I_2} \approx 1.5$~$\sqrt{\mathrm{W}}$/cm. The error bar ($\pm 15~$\%) takes into account the error on the linear fit, as well as the uncertainties on the laser waist size and on the calibration of the power meter. This calibration of $\Omega$ is used for the fit function of the PA spectra.

One notes in Fig.~\ref{fig:atintensity} that for $\sqrt{I_2}$ above 1.5~$\mathrm{W^{1/2}.cm^{-1}}$, the half splitting $\delta$ is no longer proportional to $\sqrt{I_2}$. The explanation of this effect is that the photon 2 also induces a non negligible coupling of $v=0$ with the continuum of scattering states in the $^5\Sigma_g^+$ as soon as the Rabi frequency $\Omega$ is comparable with the binding energy of the $v=14$ state of $h\times$91.35~MHz.

\subsection{Fit of the experimental spectra}
\label{fit}

For this analysis we first calculate the two S-matrix elements $S_{0,v=0}$ and $S_{0,v=14}$. $S_{0,v=14}$ and $S_{0,v=0}$ are of the same order of magnitude for the usual experimental parameters. In our fit, we use only the contribution of $S_{0,v=0}$ and neglect the contribution of $S_{0,v=14}$. The reason is that the $v=14$ molecular state is much less efficient than the $v=0$ state at heating up the cloud through its decay products. The $v=14$ molecular state decays from Penning ionization, and eventually from collisions with the surrouding gas cloud, whereas the $v=0$ molecule decays mostly radiatively, as shown in \cite{Leonardexp}. In the case of $v=14$ the decay products have a very large energy and leave immediately the magnetic trap without heating the cloud; even if they collide with a trapped atom, they expel it immediately. On the contrary, in the case of $v=0$, the molecule decays into atoms of smaller energy, a significant fraction of them can remain trapped and heat up the rest of the cloud. In addition, another argument for neglecting the contribution of $S_{0,v=14}$ is based on the analysis of the shape of the Raman signals as shown in Fig.~\ref{fig:3ramans}. The shape calculated from $S_{0,v=14}$ alone would be symmetrical, whereas the shape calculated from $S_{0,v=0}$ alone exhibits the typical asymmetrical Fano profile (see \cite{API,CCT} and references therein, and also Fig.4a in \cite{Jeroen}). Experimentally, we observe a clearly asymmetrical profile in Fig.~\ref{fig:3ramans}, which agrees well with the assumption that the heating of the cloud is predominantly due to the decay of the excited $v=0$ state.

In Fig.~\ref{fig:at}, Fig.~\ref{fig:3ramans} and Fig.~\ref{fig:3signaux}, the solid lines are the fit to the experimental data, where all parameters in equation~(\ref{eq:thav}) are known except $\gamma_{2}$. One notes that the curves fit well all the spectra, given the dispersion of the experimental points. For the spectra in each of the three figures, the fit parameters ($\Gamma$, $\Omega$, $\gamma_{1,2}$, $E_\infty$, $\Delta_{1,2}$) are exactly the same for all three signals (atom loss, peak optical density and temperature), except for an arbitrary offset and the amplitude adjustement. This fact confirms that all three signals are related to the same processes and can be indifferently exploited.

%
\section{Lifetime study of the exotic helium molecule}
\label{lifetime}

In this section, we more precisely make use of the two-photon experimental data in view of deriving the lifetime $\tau$ of the least bound state $v=14$ in the $^5\Sigma_g^+$ potential. The motivation is to obtain information on the Penning ionization within the molecular state, or in the case of atom-molecule collisions. First we show  the narrow atom-molecule dark resonance signals and discuss the precision of their theoretical fit as a function of $\gamma_{2}=1/\tau$. Then we report on linewidth measurements for the Raman peak when crucial parameters of the experiment are varied, from which we derive a value for $\tau$.

\subsection{Fit of the dark resonance signals}
\label{dark resonance}

\begin{figure}

  \includegraphics[angle=-90,scale=0.25]{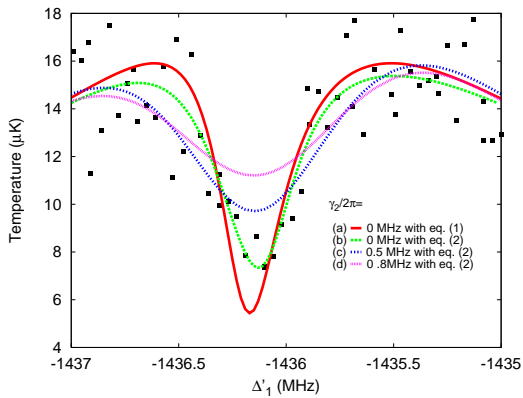}

\caption{Zoom on the atom-molecule dark resonance dip. Data are taken with $I_1 = 13$ mW/cm$^2$, $I_2 = 6.5$ mW/cm$^2$, and $\tau_{\mathrm{PA}} = 2$ ms. Data are fitted with formula (\ref{eq:s012}) for curve (a) and with formula (\ref{eq:thav}) for curves (b),(c) and (d). The value of $\gamma_2/2\pi$ is the same for curves (a) and (b). Curves (b),(c),(d) only differ from the value of $\gamma_2$. Clearly, adjustments b) and c) are both acceptable. The signal quality is here less good than in Fig. \ref{fig:3ramans}, as $I_2$ is on purpose chosen lower to avoid power broadening.
}
\label{fig:atestimation}
\end{figure}

To obtain an estimate of $\gamma_2$, we fit a typical dark resonance dip with formulas derived in section~\ref{theory}. The results are shown in Fig.~\ref{fig:atestimation}. 
First we compare theoretical curves for which $\gamma_2=0$, with and without thermal averaging, in order to figure out the influence of the temperature broadening on the dip width. In Fig.~\ref{fig:atestimation}, curve (a) neglects the thermal broadening and is just derived from formula~(\ref{eq:s012}), while curve (b) takes it into account and is derived from (\ref{eq:thav}). One notes that the result of the adjustments for curves (a) and (b) are significantly distinct. Curve (a) does not match the experimental points as well as curve (b). This shows that the thermal broadening introduces a contribution to the linewidth which cannot be ignored from the fits for the derivation of $\gamma_2$, which is coherent with the fact that $k_BT/h \approx 0.08$~MHz at the temperature $T=4$~$\mu$K at which the data of Fig~\ref{fig:atestimation} is taken.
Now we deal with curves (b),(c) and (d) of Fig.~\ref{fig:atestimation} which all take into account thermal averaging, with $\gamma_2/2\pi=0,0.5,0.8$~MHz respectively. The fit with $\gamma_2 /2\pi = 0.8$~MHz clearly does not properly match the points, whereas the values of $\gamma_2$ under 0.5~MHz give possible adjustments. This gives only a lower limit for the lifetime $\tau>0.3$~$\mu$s. We varied the temperature in the experiment and tried to derive its influence on the width of the dark resonance signal. However, the results could not be exploited due to the poor quality of the signal at low photon 2 intensity. In an attempt to derive a measurement, we then use an alternative method, based on  the linewidth of the Raman peak.

\subsection{Measurement of the lifetime with the Raman peak}
\label{raman}

We study the shape of the Raman peaks recorded as described in section ~\ref{PAsignal} and shown in Fig.~\ref{fig:3ramans}. Here we cannot apply a fit function to these Raman signals as was done with the dark resonance dip, because experimental conditions do not allow it. Actually Raman signals with a good signal to noise ratio are taken at large $\Delta_1-\Delta_2$ detunings (from 10~MHz to above 20~MHz), using  large intensities and/or long photoassociation pulses: it is thus impossible to simultaneously record the main PA line with the same experimental parameters as those requested for the Raman peak, because such high laser power blows up the cloud at resonance when $\Delta_1=0$. Without a full spectrum, some parameters of the fit are missing (for instance the position of the main peak which may be light shifted) and the fit cannot be exploited. Instead we just measure the full width at half maximum (FWHM) of the Raman peak, in order to extract values of $\gamma_2$ after eliminating the broadening due to the main experimental parameters: intensity of the second photon, PA pulse duration, temperature of the atomic sample.

\subsubsection{Line broadening with the PA pulse parameters}
\label{intensity}

In all these experiments, the first photon is used as a probe. Its intensity is kept constant and as weak as possible, in order to avoid any alteration of the spectra. We carefully checked that it does not introduce any broadening of the Raman peaks by varying $I_1$ much above the value necessary to record the spectra. In addition we also calculated that the value of the transition rate $\Gamma$ has no influence on the Raman peak width under the experimental conditions ($\Delta_1-\Delta_2=10$~MHz).   

\begin{figure}

  \includegraphics[angle=-90,scale=0.25]{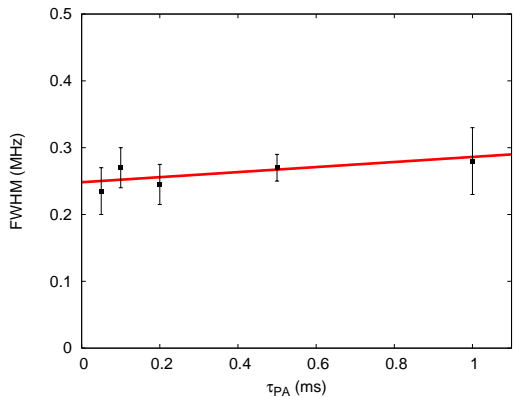}

\caption{FWHM of the Raman peak versus the PA pulse duration $\tau_{PA}$. All other experimental parameters are fixed : $I_1 = 650$ mW/cm$^2$,~$I_2 = 65$ mW/cm$^2$, $\Delta_1-\Delta_2 = 20$ MHz, $T \approx 4$ $\mu$K.
}
\label{fig:PAduration}
\end{figure}

The PA pulse duration $\tau_{\mathrm{PA}}$ is usually adjusted according to the laser intensity $I_2$ of the second photon, in order to keep the temperature raise signal observable with the best possible signal to noise ratio. 
The PA pulse duration may broaden the line for two reasons. On the one hand, if it is too small, the linewidth would be Fourier-limited, which is not the case for the shortest pulse we used (50 $\mu s$). On the other hand if it is too long, the  line could be distorted, as the number of atoms decreases all along the pulse duration, all the more that the laser frequency is set close to resonance. To check if there is an influence of the pulse duration $\tau_{\mathrm{PA}}$, we varied it from 50~$\mu$s to 1~ms, all other parameters being identical. The result is shown in figure \ref{fig:PAduration}. The straight line is the fit to the experimental data, which is compatible with no variation of the linewidth with $\tau_{PA}$.

Finally we checked the influence of the intensity $I_2$ of the second photon. Actually for recording good Raman spectra such as those in Fig.~\ref{fig:3ramans}, the $I_2$ values are much larger than for recording the Autler-Townes signals in  Fig.~\ref{fig:at}. The larger the $\Delta_1-\Delta_2$ detuning, the larger the necessary $I_2$ intensity. For instance, one requests $I_2$ values about 200 times larger for $\Delta_1-\Delta_2=20$~MHz than for $\Delta_1-\Delta_2=0$. On the other hand, the theory shows that the broadening of the Raman peak decreases with $\Delta_1-\Delta_2$. There is thus a compromise to find for the choice of the appropriate detuning limiting the $I_2$ intensity and the subsequent broadening of the Raman peak. 

\begin{figure}

  \includegraphics[angle=-90,scale=0.25]{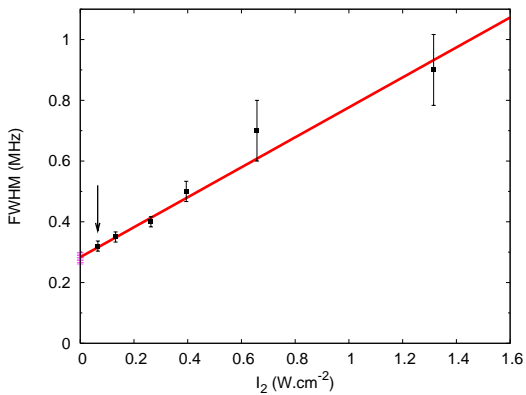}

\caption{FWHM of the Raman peak versus intensity $I_2$ of the second laser. The experimental parameters are fixed to $I_1 = 660$ mW/cm$^2$, $\Delta_1-\Delta_2 = 20$ MHz, $T \approx 4$ $\mu$K.
}
\label{fig:Rintensity}
\end{figure}

We changed the $I_2$ intensity  of the second photon within a range compatible with a clearly visible Raman peak. For each $I_2$ value we set the appropriate PA pulse duration $\tau_{PA}$, ranging from 10 to 200~$\mu$s for $I_2$ values from 0.1 to 1.4 W/cm$^2$. Note that the change of pulse duration has no influence on the width of the Raman peak, as discussed above.  We observe a broadening of the measured linewidth with the laser intensity $I_2$, as shown in Fig.~\ref{fig:Rintensity}, which is compatible with a linear fit.  Experimental error bars are larger at high intensity, as the signal to noise progressively degrades for $I_2$ above 0.5~$\mathrm{W.cm^{-2}}$. From this we conclude that it is important to operate at $I_2$ intensities as weak as possible if one aims at measuring $\gamma_2$. One finds $(0.28 \pm 0.02)$~MHz for the FWHM value extrapolated at zero intensity, where the error bar is twice the one given by standard linear fit procedure \cite{NR}. 

\subsubsection{Line broadening with temperature}
\label{temperature}

\begin{figure}

  \includegraphics[angle=-90,scale=0.25]{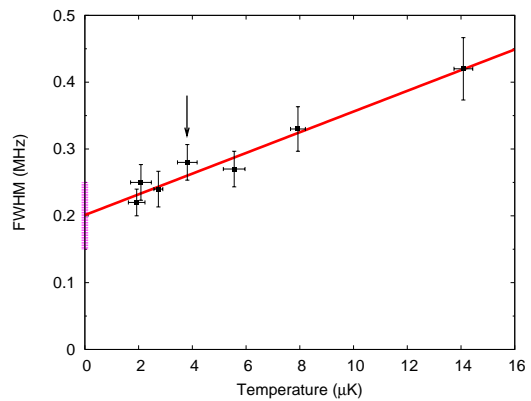}

\caption{FWHM of the Raman peak versus temperature of the atomic cloud. The temperature is changed by ajusting the evaporative cooling stage. The experimental parameters are fixed to $I_1 = 660$ mW/cm$^2$, $I_2 = 66$ mW/cm$^2$, $\tau_{\mathrm{PA}} = 200$ $\mu$s, $\Delta_1-\Delta_2 = 20$ MHz.
}
\label{fig:temperature}
\end{figure}

The most important contribution to the linewidth at low laser intensity is the thermal broadening resulting from the energy distribution of the atoms, as already understood from section~\ref{dark resonance} on the atom-molecule dark resonance.  Fig.~\ref{fig:temperature} shows the Raman peak linewidth measured as a function of the average temperature of the gas. All the points are taken for the same $I_2$ intensity, chosen as low as possible, of order $0.1$~W/cm$^2$, for which the intensity broadening is not significant (see the arrow 
in Fig.~\ref{fig:Rintensity}). For a temperature of 4~$\mu$K at which the data of Fig.~\ref{fig:Rintensity} are taken, the linear fit of Fig.~\ref{fig:temperature} gives a FWHM value of (0.25~$\pm$~0.03)~MHz, where the error bar is defined similarly to the one of Fig.~\ref{fig:Rintensity}. It is in good agreement with the value of (0.28~$\pm$~0.02)~MHz found at nearly zero $I_2$ intensity in Fig.~\ref{fig:Rintensity}. There is a consistency between the results for the linewidth studied as a function of the intensity and of the temperature.

The results plotted in Fig.~\ref{fig:temperature} give a FWHM value at zero temperature of $(0.20 \pm 0.05)$~MHz, taking an error bar twice as large as that given by a linear fit of the data. 

One can also deduce from the fit that the thermal broadening contribution to the linewidth at the experimental temperature of 4~$\mu$K (see the arrow in Fig.~\ref{fig:temperature}) is $(0.05 \pm 0.02)$~MHz.

\subsubsection{\label{sec:lifevalue}Value of the lifetime}
\begin{figure}

  \includegraphics[angle=-90,scale=0.25]{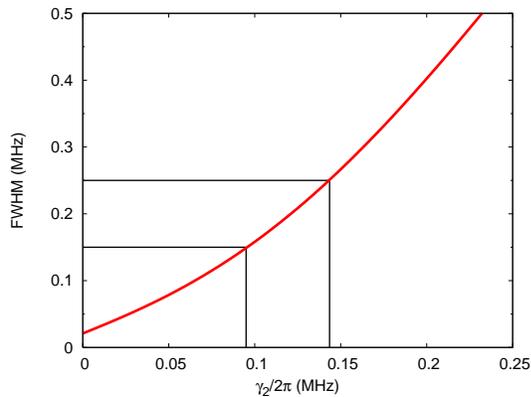}

\caption{FWHM of the Raman peak versus $\gamma_2$. This curve is calculated at zero temperature for the typical intensity (see arrow in Fig.~ \ref{fig:Rintensity}) and detuning (20 MHz) used in the experiment. For $\gamma_2=0$, the small remaining width is due to power broadening.}
\label{fig:Rgamma2}
\end{figure}

As a final result for the FWHM of the Raman signal, we adopt for the value deduced from the results of Fig.~\ref{fig:temperature} at zero temperature :
$$FWHM=(0.20\pm0.05)~\mathrm{MHz}$$
The value of the FWHM can be related to those of parameter $\gamma_2/2\pi$. The relation between the two can be calculated as shown in Fig~\ref{fig:Rgamma2} using references \cite{Bohn96,Bohn99}. We thus find:
$$\gamma_2/2\pi=(0.12\pm0.025)~\mathrm{MHz}$$
In conclusion, we derive a final value for the lifetime $\tau=1/\gamma_2=(1.4\pm0.3)~\mu\mathrm{s}$. The value of $\tau$ found here with the Raman spectra does not contradict the lower limit given in section~\ref{dark resonance} as derived from the dark resonance signals. It also agrees with the measurement reported in our previous publication \cite{2photon}. Let us stress the point that the present value of $\tau$ is the lifetime of the molecules as measured in the presence of the atomic cloud.

\section{Limitation of the lifetime by collisions}
The most probable decay process of the exotic helium molecule in the $v=14$ state is Penning ionization. Another possible cause of lifetime shortening are the collisions of the molecules with the atoms of the cloud or other molecules : the inelastic collisions with the surrounding medium could cause a broadening of the Raman peak, as observed in the case of Rb$_2$ molecules in a condensate \cite{Wynar,Grimm}. The number of molecules produced is much smaller than the number of atoms, so that the contribution of molecule-molecule collisions to the limitation of the lifetime is negligible compared to the contribution of atom-molecule collisions even if one assumes an inelastic molecule-molecule rate at the unitarity limit. In order to investigate the rate of the latter process, we performed some measurements with different atom densities (section \ref{density}), and a different molecular electronic state (section \ref{bubu})

\subsection{Influence of the atom density}
\label{density}

\begin{figure}

  \includegraphics[angle=-90,scale=0.25]{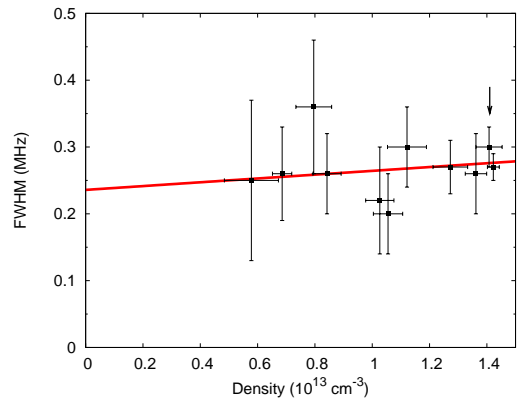}

\caption{FWHM of the Raman peak versus atomic cloud density. The density is varied by adjusting the evaporative cooling stage. The experimental parameters are fixed to $I_1 = 660$ mW/cm$^2$, $I_2 = 66$ mW/cm$^2$, $\tau_{\mathrm{PA}} = 500$ $\mu$s, $\Delta_1-\Delta_2 = 20$ MHz.}
\label{fig:densite}
\end{figure}

The measured linewidths in section \ref{lifetime} are taken at an atom density of $1.4~10^{13}~$cm$^{-3}$. It provides an upper limit for the inelastic atom-molecule collision rate $K_{inel}$ of $6~10^{-8}\mathrm{cm^3.s^{-1}}$. Such a value for $K_{inel}$ is unlikely, as it is very large. Similar experiments with weakly-bound $Rb_2$ molecules in the electronic ground state gave $K_{inel}<8~10^{-11}\mathrm{cm^3.s^{-1}}$ \cite{Wynar}.  We varied the density in our experiment in order to verify that the influence of collisions is negligible. The atom density range for our measurement is limited if we want to keep the same experimental parameters at all densities, in particular the intensity $I_2$ : at too low atomic density (below $n_0 =0.6 \times 10 ^{13}$cm$^{-3}$), signals are blowned up by the PA pulse; conversely we reach a maximum density of $n_0 =1.5 \times 10 ^{13}$cm$^{-3}$ with the characteristics of our trapping procedure. Our results for the FWHM linewidth of the Raman peaks as a function of the atomic density are plotted in Fig.~\ref{fig:densite} for a low $I_2$ intensity and a temperature of 4~$\mu$K. The error bars are rather large at low density, due to the relatively low signal-to-noise ratio. The straight line in Fig.~\ref{fig:densite} is a linear fit to the data, corresponding to a FWHM at zero density of (0.24$\pm$0.08)~MHz and a broadening of (0.029$\pm$0.057)~MHz/$10^{13}$cm$^{-3}$, which is compatible with zero. 

\begin{figure}[h]
\resizebox{0.8\columnwidth}{!}{%
  \includegraphics{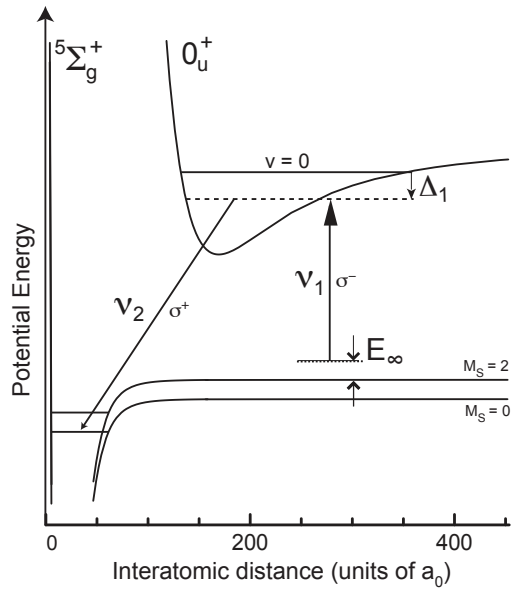}
}
\caption{Levels involved in the two-photon photoassociation experiment when the polarization of the bound-bound photon 2 is $\sigma^+$. Energies are not to scale
}
\label{fig:potplus}
\end{figure}

\subsection{Influence of the electronic state of the molecule}
\label{bubu}

We also performed a few measurements with a $\sigma^+$ polarization for photon 2, which drives a bound bound transition to the $v=14$ molecular state in a $^5\Sigma_g^+$ potential connected to the $M_S=0$ asymptote, where $M_S$ is the projection of the total spin of the pair of atoms on the lab-fixed quantization axis (see Fig. \ref{fig:potplus}). In this case, the frequency of photon 2 is shifted by $4\mu_B B\approx20$~MHz compared to the case of Fig. \ref{fig:pot}

\begin{figure}[htbp]

 \includegraphics[angle=-90,scale=0.25]{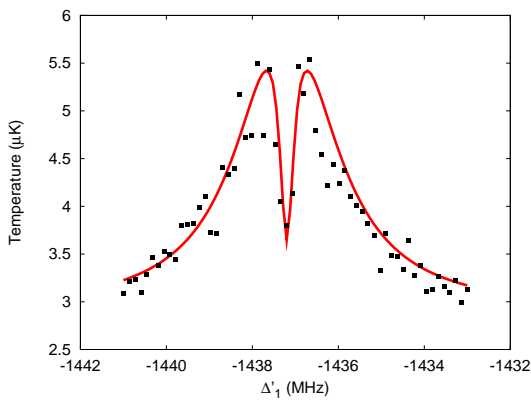}

\caption{Atom-molecule dark resonance in two-color photoassociation with a $\sigma^+$ polarized photon 2, according to the scheme of Fig. \ref{fig:potplus}.
}
\label{fig:DRsigp}
\end{figure}

We do not observe a larger width for the signals in this case, as shown in Fig. \ref{fig:DRsigp}. This is another argument demonstrating that there is no atom-molecule collision contribution in the measured linewidth of section \ref{sec:lifevalue}. Actually we expect the inelastic atom-molecule rate to be much larger in the case when the unexcited molecule is in the $M_S=0$ state of the $^5\Sigma_g^+$ potential, than in the case $M_S=2$. The reason is that the inhibition by the spin conservation during the atom-molecule collision occurs for the molecule in the $M_S=2$ state, but not in the $M_S=0$ state. This argument is based on similar spin conservation consideration, as for the inhibition of Penning ionization by spin polarization in atomic collisions \cite{Fedichev}.

Finally, we conclude from these studies that the linewidth of the PA signals cannot be attributed to atom-molecule collisions.

%
\section{Calculation of spin-dipole induced lifetime of the molecule in the $v=14$ state  }
\label{calculation}

Ionization processes limit the lifetime of the exotic molecule in the least-bound vibrational $v=14$ state. Estimates of such mechanisms are here given for comparison with our measurement. As the two $2^3S_1$ metastable helium atoms are spin-polarized when they form the $v=14$ molecule in the $^5\Sigma_g^+$ potential, Penning ionization and associative ionization are inhibited by the electronic spin polarization, unless spin relaxation takes place. Such relaxation process result from the weak spin-dipole coupling between the quintet $^5\Sigma_g^+$ state and the singlet $^1\Sigma_g^+$ state \cite{Fedichev}\cite{Venturi}. At ultralow temperature and weak magnetic field, relaxation-induced ionization is the leading decay channel \cite{Fedichev}. 

Recently a detailed calculation of the spin-dipole induced lifetime of the least bound $^5\Sigma_g^+$ state of metastable helium was produced by I.Whittingham's team \cite{Whitt06}. The spin-dipole interaction between the $2^3S_1$ metastable atoms couples the spin polarized S$M_{S}=|22\rangle$ state to the $|21\rangle$, $|20\rangle$ and $|00\rangle$ electronic spin states. As Penning ionization and associative ionization from the $S = 0$ singlet state are highly probable, the spin-dipole induced coupling results in a finite lifetime of the $|22\rangle$ state. The spin-dipole is treated as a perturbation and the method presented in \cite{Whitt06} is an adaptation of those given in \cite{Venturi}. They use the ab initio calculations of \cite{Prz05} for the $^5\Sigma_g^+$ potential and an optical potential in order to model Penning ionization in the singlet channel. The coupling between the initial quintet state and the singlet state produces a complex energy shift, the imaginary part of which provides the lifetime of the least bound state of the $^5\Sigma_g^+$ potential.  They find values between $132.1~\mu$s and $177.9~\mu$s for the lifetime.

Our group recently undertook another calculation for the same process. It is based on \cite{Goran,Fedichev}. The Penning ionization process is modelled using a fully absorbing wall at short distance in the $S = 0$ singlet interaction potential. One calculates the flux of atoms absorbed on such a wall to derive a Penning ionization rate. This calculation gives a value of the order of $120~\mu$s for the spin-dipole induced lifetime of the doubly excited molecule in the $v=14$ state \cite{Gora}. Both theoretical values of $\tau$ provide the same order of magnitude.

%
\section{Conclusion}
\label{conclusion}

In this article we fully explained the origin of all signals recorded in two-photon photoassociation experiments operated in an ultracold gas of metastable helium atoms. The atom-molecule dark resonance, as well as the Raman signals, were used to provide information on the exotic molecule in the least bound vibrational state of the $^5\Sigma_g^+$ potential of a pair of spin polarized $2^3S_1$ metastable atoms. A study of the lifetime of this molecule could be derived from the linewidth measurements of the spectra. We find the value $\tau=1.4\pm0.3~\mu$s for the exotic molecule. It cannot be attributed to collisions with the atomic cloud environment. It is neither explained by theoretical estimates of the rate of Penning ionization inside the molecule. The discrepancy is still to be understood. If one excludes an experimental artefact, one can think of either a not fully justified assumption in the Penning ionization calculation either another non explored mechanism. 

 It would be interesting to continue these photoassociation experiments in conditions more appropriate to derive a more precise experimental value of the intrinsic lifetime of the molecule. We plan to add a channel electron multiplier to our setup in order to monitor the ion production rate. It would help understanding the decay mechanism of the molecular $v=14$ state, knowing that the molecular state in the $0_u^+$ electronic state does not decay through Penning ionization \cite{Rijnbach}. Furthermore, our group plans to perform new two-photon photoassociation experiments turning to the Bose-Einstein condensed phase rather than the ultracold gas. By operating at much lower temperature one could more easily disentangle the thermal broadening from the intrinsic lifetime of the molecular state. The comparison with the two recent theoretical estimates would be fruitful. 
 
We thank C. Cohen-Tannoudji, G. Shlyapnikov and E. Arimondo for very helpful discussions.
%
%
%

%
%

\begin{thebibliography}{}
%
\bibitem{Whitt06}
T. J. Beams, G. Peach,I. B. Whittingham, Phys. Rev. A \textbf{74}, (2006) 014702
%
\bibitem{Gora}
M. Portier and G. V. Shlyapnikov, in preparation
%
\bibitem{Goran}
G. V. Shlyapnikov, J. T. M. Walraven, U. M. Rahmanov, M. W. Reynolds, Phys. Rev.
Lett. \textbf{73}, (1994) 3247.
%
\bibitem{Fedichev}
P. O. Fedichev, M.W. Reynolds, U. M. Rahmanov, and
G.V. Shlyapnikov, Phys. Rev. A \textbf{53}, (1996) 1447.
%
\bibitem{Venturi}
V. Venturi,I. B. Whittingham, P. J. Leo, G. Peach, Pys. Rev. A \textbf{60}, (1999) 4635.
%
\bibitem{Robert}
A. Robert, O. Sirjean, A. Browaeys, J. Poupard, S. Nowak,
D. Boiron, C. I. Westbrook, and A. Aspect, Science \textbf{292}, (2001)
461.
%
\bibitem{Pereira}
F. Pereira Dos Santos, J. Le´onard, J. Wang, C. J. Barrelet,
F. Perales, E. Rasel, C. S. Unnikrishnan, M. Leduc,
and C. Cohen-Tannoudji, Phys. Rev. Lett. \textbf{86},
(2001) 3459.
%
\bibitem{2photon}
S. Moal, M. Portier, J. Kim, J. Dugu\'e , U.D. Rapol, M. Leduc, and C. Cohen-Tannoudji, Phys. Rev. Lett. \textbf{96}, (2006) 023203.
%
\bibitem{Bohn96}
J. L. Bohn and P. S. Julienne, Phys. Rev. A \textbf{54}, (1996) R4637.
%
\bibitem{Bohn99}
J. L. Bohn and P. S. Julienne, Phys. Rev. A \textbf{60}, (1999) 414.
%
\bibitem{Leonardexp}
J. L\'eonard, M.Walhout, A. P. Mosk, T. M\"uller, M. Leduc, and C. Cohen-Tannoudji, Phys. Rev. Lett. \textbf{91}, (2003) 073203.
%
\bibitem{Kim}
J. Kim, U.D. Rapol, S. Moal, J. L´eonard, M. Walhout, and M. Leduc, Eur. Phys. J. D \textbf{31}, (2004) 227-237.
%
\bibitem{API}
C. Cohen-Tannoudji, J. Dupont-Roc, and G. Grynberg, \emph{Atom-
Photon Interactions}, (Wiley, New York, 1998).
%
\bibitem{Fano61}
U. Fano, Phys. Rev. \textbf{124}, (1961) 1866.
%
\bibitem{Napolitano}
R. Napolitano, J. Weiner, C.J. Williams, P.S. Julienne,
Phys. Rev. Lett. \textbf{73}, (1994) 1352.
%
\bibitem{CCT}
C. Cohen-Tannoudji and B. Lounis, J. Phys. II (France) \textbf{2}, (1992) 579.
%
\bibitem{Jeroen}
J. C. J. Koelemeij and M. Leduc, Eur. Phys. J. D \textbf{31}, (2004) 263.
%
\bibitem{NR}
Numerical Recipes in FORTRAN, Cambridge University Press
%
\bibitem{Wynar}
R. Wynar, R. S. Freeland, D. J. Han, C. Ryu, and D. J. Heinzen, Science \textbf{287}, (2000) 1016.
\bibitem{Grimm}
K. Winkler, G. Thalhammer, M. Theis, H. Ritsch, R. Grimm, and J. Hecker Denschlag, Phys. Rev. Lett. \textbf{95}, (2005) 063202.
%
%
\bibitem{Prz05}
M.Przybytek, B.Jeziorski, J.Chem.Phys. \textbf{123}, (2005) 134315.
\bibitem{Rijnbach}
M. Rijnbach, Dynamical spectroscopy of transient $He_2$ molecules, PhD Thesis, Universiteit Utrecht (2004) http://igitur-archive.library.uu.nl/dissertations/2004-0120-094105/inhoud.htm



\end{thebibliography}
\end{document}